\documentclass[prl,twocolumn,showpacs,amssymb,amsfonts,amsmath]{revtex4}
\usepackage{mathrsfs,bbm}
\usepackage{graphicx}

\def\textbf#1{{\bf #1}}
\def\be{\begin{equation}}
\def\ee{\end{equation}}
\def\ben{\begin{eqnarray}}
\def\een{\end{eqnarray}}
\def\eea{\end{array}}
\def\bea{\begin{array}}
\newcommand{\bei}{\begin{itemize}}
\newcommand{\eei}{\end{itemize}}

\def\blacksquare{\vrule height 4pt width 3pt depth2pt}

\newcommand{\srn}[2]{\big\langle\hspace{-0.1cm}\big\langle #1 \big\rangle\hspace{-0.1cm}\big\rangle_{#2}}
\newcommand{\Tr}[0]{\mathrm{\mbox{Tr}}}
\newcommand{\ot}{\otimes}
\newcommand{\ket}[1]{|#1\rangle}
\newcommand{\bra}[1]{\langle#1|}
\newcommand{\proj}[1]{\ket{#1}\bra{#1}}

\begin{document}

\title{Universal observable detecting all two--qubit entanglement and determinant based separability tests }

\author{Remigiusz Augusiak}\email{remik@mif.pg.gda.pl}
\author{Maciej Demianowicz}
\author{Pawe\l{} Horodecki}

\address{Faculty of Applied Physics and Mathematics,
Gda\'nsk University of Technology, 80--952 Gda\'nsk,
Poland}

\begin{abstract}
We construct a single observable measurement of which mean value
on four copies of an {\it unknown} two-qubit state is sufficient
for unambiguous decision whether the state is separable or
entangled. In other words, there exists a universal collective
entanglement witness detecting all two-qubit entanglement. The
test is directly linked to a function which characterizes to some
extent the entanglement quantitatively. This function is an
entanglement monotone under so--called local pure operations and
classical communication (pLOCC) which preserve local dimensions.
Moreover it provides tight upper and lower bounds for negativity
and concurrence. Elementary quantum computing device estimating
unknown two-qubit entanglement is designed.

\end{abstract}

\maketitle


{\it Introduction .-} One of the main challenges of both
theoretical and experimental Quantum Information Theory is a
determination of entanglement properties of a given state. There
is an extensive literature covering the problem of deciding
entanglement of a state
\cite{QuEntEPR,sep,Terhal,Developed,Experiments,UncertaintyCV,UncertaintyDisc}.
As one knows from the seminal paper of Peres and Wootters
\cite{PeresWotters} collective measurement on several copies of a
system in a given quantum state may provide better results than
measurements performed on each copy separately. This fact was
reflected in the method of  entanglement detection with collective
measurements. The method initiated for pure states
\cite{SH00,Acin}, then developed for mixed states with help of
quantum networks
\cite{PHAE,PHPRL,Carteret,Noiseless,Estimator,Filip,Fiurasek} and
the concept of collective entanglement witnesses \cite{Witnesses},
has found its first experimental demonstration in coalescence-anti
coalescence coincidence experiment \cite{NonlinExp}. In
particular, somewhat surprisingly, it was shown how to estimate
and/or even measure amount of entanglement (concurrence) without
prior state reconstruction \cite{PHAE,PHPRL,Carteret}. Recently
the method got the new twist thanks to application of such
collective measurements \cite{ConcurrenceDet,Magda,Mintert,Aolita}
that are directly related to quantum concurrence (see
\cite{MiKuBu2005}) including photon polarization-momentum
experimental demonstration for pure states in distant laboratories
paradigm \cite{ConcurrenceDet}. Recently collective entanglement
witnesses were also shown to lead to easily measurable lower
bounds on entanglement \cite{Mintert}. The idea of collective
entanglement witnesses was also implemented in continuous
variables setup \cite{Magda}.

We show that a single observable if measured on four copies of a
unknown two--qubit state is sufficient for discrimination between
entanglement and separability of it. Moreover it can serve for
limited quantitative purposes. To this aim we explore the
two--qubit separability test (equivalent to the PPT one
\cite{sep,Peres}) stating that a state is separable iff the
determinant of its partially transposed density matrix is
nonnegative \cite{Sanpera,Verstraete}. The result, known for a few
years, was barely mentioned in the literature in that form (see
{\it e.g.} \cite{Slater}) and up to our knowledge this is the
first time an operative physical meaning is assigned to it. Namely
we introduce a state function, straightforwardly connected to the
test, which is a monotone under pLOCC with fixed dimensions (see
\cite{VerstraeteSLOCC,Gour}) and  only single collective
observable is enough to measure it experimentally, and provides
tight upper and lower bounds for the two-qubit negativity and
concurrence.

Further we discuss how the result allows to build a small quantum
device implementing a kind of elementary algorithm, namely,
detecting entanglement in an unknown two--qubit state. Our method
has a significant advantage over prior methods
\cite{PHPRL,Carteret} as we require only one  collective
measurement. In comparison to the result of Ref. \cite{Mintert},
where a single observable provides a concurrence lower bound which
sometimes is not conclusive, we achieved sharp test which is to
some extent quantitative.

We also discuss higher dimensional and multiparty generalizations.
In particular, we find that reduction criterion \cite{xor,red} on
composite $2 \otimes d$ systems with the map applied to the second
subsystem is equivalent to a single determinant condition and as
such can be checked via measurement of a single observable.

{\it The criterion.-} Here we discuss the necessary and sufficient
condition for two-qubit separability in terms of a determinant of
a partially transposed density matrix. The observation follows
from the facts from papers of Sanpera {\it et al.} \cite{Sanpera}
and Verstraete {\it et al.} \cite{Verstraete}. Here we prove more
general statement about the reduction criterion, exploiting its
equivalence to PPT test on two qubits. Let us consider the
reduction map defined as $\Lambda_{r}(A)=\Tr(A)\mathbbm{1}_{d}-A$
on any $d\times d$ matrix $A$ with $\mathbbm{1}_{d}$ standing for
an identity acting on $\mathbb{C}^{d}$. The following proposition
holds.

{\it Proposition 1.-} For any $2 \otimes d$ state  $\varrho$ the
reduction criterion with respect to the system $B$ is satisfied
iff
\begin{equation}
\det\{[I\otimes \Lambda_{r}](\varrho)\} \geq 0 \label{negative}.
\end{equation}
In particular any two-qubit state is separable iff
\begin{equation}
\det\varrho^{\Gamma}\geq 0. \label{wyznacznik}
\end{equation}

{\it Proof.} The necessity of the condition  is obvious. Let us
prove sufficiency. To this aim we may assume that our $2 \otimes
d$ state $\varrho$ has nonsingular reduced density matrix
$\varrho_{A}=\Tr_{B}\varrho$, as otherwise it would be a product
state. Applying a local filter
$\mathscr{V}_{A}=(\varrho_{A}^{-1}/2)^{1/2}$ and utilizing
previous observation, one obtains $\det\{[I\otimes
\Lambda_{r}](\varrho)= [\det(\varrho_{A}/2)]^{2} \det \{[I\otimes
\Lambda_{r}](\tilde{\varrho})\}$,
where the state $\tilde{\varrho}$ is a result of local filtering.
Now there is an immediate observation that for any positive
$\Lambda$  positivity of  $[I\otimes \Lambda](\varrho)$ is {\it
equivalent} to the positivity of the new state being the result of
local filtering on system $A$ with arbitrary nonsingular filter
$\mathscr{V}_{A}$. Since we deal with the nonsingular
$\mathscr{V}_{A}$, the original state $\varrho$ violates the
reduction criterion iff the state  $\tilde{\varrho}$ does. Suppose
this is the case. Since  the first subsystem of the latter is in a
maximally mixed state, i.e.,
$\tilde{\varrho}_{A}=(1/2)\mathbbm{1}_{2}$ one easily infers (cf.
\cite{Estimator}) that in order to violate the criterion
$\tilde{\varrho}$ {\it must}  have one eigenvalue that is greater
than one-half. Then the operator $[I\otimes
\Lambda_{r}](\tilde{\varrho})=(1/2)\mathbbm{1}_{2d}-\tilde{\varrho}$
clearly has the spectrum with all nonzero values in which only one
is negative. This finally gives $\det\{[I\otimes
\Lambda_{r}](\tilde{\varrho})\}<0$ which (as we already mentioned)
is equivalent to $\det\{[I\otimes \Lambda_{r}](\varrho)\}<0$. Thus
violation of reduction criterion by $2 \otimes d$ state on the
second subsystem is equivalent to violation of (\ref{negative}).

To prove the second part, we only need to observe that
$\det\varrho^{\Gamma}=\det(\mathbbm{1}_{2}\otimes\sigma_{y}
\varrho^{\Gamma}\mathbbm{1}_{2}\otimes\sigma_{y}) =\det\{[I\otimes
\Lambda_{r}](\varrho)\}$ and recall that reduction criterion is
equivalent to PPT test on two-qubit states. This concludes the
proof.\blacksquare

{\it Quantifying entanglement.-} A question important from an
experimental point of view is whether a a function of determinant
of a partially transposed density matrix can serve for {\it
quantitative} purposes. We obtain partial positive answer.

First we introduce the function defined on $d\otimes d$ states
\begin{eqnarray}
\label{miara} \pi_{d}(\varrho)=
\left\{%
\begin{array}{ll}
    0, & \quad\hbox{$\det\varrho^{\Gamma}\ge 0$;} \\
    d\sqrt[2d]{\displaystyle\left|\det\varrho^{\Gamma}\right|}, & \quad\hbox{$\det\varrho^{\Gamma}<0$.} \\
\end{array}
\right.
\end{eqnarray}
Let us observe that $\pi_{d}(|\psi\rangle)= d |\det
A^{\psi}|^{2/d}$ for any pure state
$\ket{\psi}=\sum_{i,j}A^{\psi}_{i,j} |i\rangle |j \rangle$. This
leads to the fact that $\pi_{d}(\ket{\psi})=G_{d}(\ket{\psi})$,
where $G_{d}$ is called $G$-concurrence and is defined as a,
scaled by the dimension factor, geometric mean value of Schmidt
numbers (see \cite{Sinolecka,Gour}). The latter is known to be a
monotone under LOCC not changing dimensions of the state, and as
such is considered as an entanglement measure
\cite{VerstraeteSLOCC,Sinolecka,Gour}. Below we prove that
$\pi_{d}$ satisfies monotonicity property under some restricted
class of LOCC (invariance under local unitary operations is
obvious due to properties of determinant), namely the ones for
which local operations are pure in a sense they consist only of
single Kraus operators. We call them pure LOCC (pLOCC). To this
aim let us assume that $\varrho$ is entangled. Then we have the
following.

{\it Proposition 2.-} For any pLOCC not changing the dimension of
a state, which transform initial state $\varrho$ to
$\varrho^{(i)}$ with probability $p_i$  the following holds
\begin{equation}
\sum_{i}p_{i}\pi_{d}(\varrho^{(i)})\leq \pi_{d}(\varrho).
\end{equation}
\indent{\it Proof.} Reasoning from \cite{Negativity} (the measure
is symmetric under the change of particles) allows to restrict
ourselves to only single measurement on Bob's side. These are
described by the family of completely positive operators
$\mathcal{M}_i$ with single Kraus decomposition (we consider only
pLOCC) i.e. their action is as follows
$\mathcal{M}_{i}(\varrho)=\mathbbm{1}_{d}\ot
M_{i}\varrho\mathbbm{1}_{d}\ot M_{i}^{\dagger}$. We take square
$M_{i}$ ($\sum_i M_{i}^{\dagger}M_{i}\le \mathbbm{1}_d$) to
fulfill the requirement of not changing the dimension. Since
$[\mathcal{M}_{i}(\varrho)]^{\Gamma
_A}=\mathcal{M}_{i}(\varrho^{\Gamma _A})$, we have
\begin{eqnarray}
&&\hspace{-0.4cm} \sum_{i}p_{i}\pi_{d}(\varrho^{(i)})  =
  d\sum_{i}p_{i}\sqrt[2d]{|\det(1/p_i)\left(\mathbbm{1}_{d}\ot M_{i}\varrho
  \mathbbm{1}_{d}\ot M_{i}^{\dagger}\right)^{\Gamma}|}\nonumber\\
   &&\hspace{-0.2cm}= \sum_{i}\sqrt[2d]{\det(\mathbbm{1}_{d}\ot
   M_{i}^{\dagger}M_{i})}\pi_{d}(\varrho)
   \leq
   \sqrt[d]{\det\sum_i M_{i}^{\dagger}M_{i}}\pi_{d}(\varrho)\nonumber
\end{eqnarray}
where the last inequality follows from Minkowski determinant
theorem. Now taking into account normalization condition for $M_i$
we conclude that the last term is less or equal to $\pi _d$
finishing the proof. \blacksquare

Unfortunately $\pi _d$ is not a general LOCC monotone. This can be
shown by performing twirling on entangled Bell diagonal
states, which in general increases $\pi _d$.

Let us now focus on the two--qubit states. Below we will establish
a connection of $\pi_{2}$ with concurrence $C$ and negativity $N$
\cite{footnote}. As shown in Ref. \cite{Verstraete2}, the
concurrence of a density matrix transformed with a filter
$A\otimes B$ changes by the factor $|\det
AB|/\mathrm{Tr}(AA^{\dagger}\otimes BB^{\dagger}\varrho)$. As it
turns out $\pi _2$ of the state transformed in this way changes
identically. Moreover the filters are known to be sufficient for
transformation of any non--singular two--qubit state to a
Bell--diagonal one \cite{Belldiag}. It is then enough to check the
relation between $C$ and $\pi_{2}$ for these states. Taking the
entangled state $\varrho$ to be the mixture of Bell states with
probabilities $\{p_i\}_{i=1}^4$ we obtain
$\pi_2(\varrho)=\Pi_i\sqrt[4]{|1-2p_i|}$, which with an assumption
$p_1\ge p_i$, gives $\pi_2(\varrho)\ge 2p_1-1$. This however means
that $\pi_2$ is bounded from below by $C$ as for Bell diagonal
states it is just equal to rhs of the above. Obviously $\pi_2$
provides also an upper bound for negativity as the latter is
always less or equal to $C$ \cite{Verstraete}:
\begin{equation}\label{upper}
N(\varrho)\le C(\varrho)\le \pi_2(\varrho).
\end{equation}
One may also provide tight lower bound on $N(\varrho)$ and
$C(\varrho)$ in terms of $\pi_{2}(\varrho)$. To this aim notice
that $\pi_2(\varrho)=2\sqrt[4]{(1/2)N(\varrho)\lambda_1
\lambda_2\lambda_3}$, where $\lambda_i$ are the positive
eigenvalues of $\varrho^{\Gamma}$. Their product is maximal when
they are equally distributed. This observation with the aid of the
fact that $\sum_{i=1}^3 \lambda_i -N/2=1$ lead us to
\begin{equation}\label{lower}
\pi_2 (\varrho)\le
\sqrt[4]{N(\varrho)\left(\frac{N(\varrho)+2}{3}\right)^3}\le
\sqrt[4]{C(\varrho)\left(\frac{C(\varrho)+2}{3}\right)^3}.
\end{equation}

In conclusion, $\pi_{2}$ although being not full entanglement
monotone quantifies all the two-qubit entanglement in a nontrivial
way providing tight lower and upper bounds for other entanglement
measures (see Fig.1.).

\begin{figure}\label{Fig1}
\includegraphics[width=6cm]{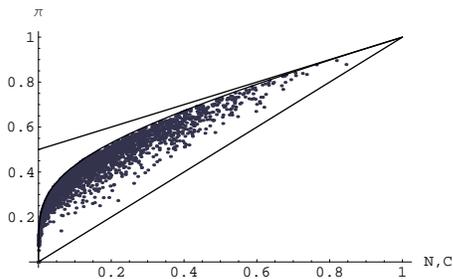}
\caption{Plot of $\pi_2$ {\it versus} $N$ (or $C$) for randomly
generated density matrices with bounds obtained in (\ref{upper})
and (\ref{lower}). We have also added a bound $\pi_2\le
(1/2)(N+1)$ obtained from geometric-arithmetic inequality applied
to the absolute values of eigenvalues of a partially transposed
density matrix. The reader is encouraged to consult
\cite{MiranGrudka}.}
\end{figure}

It can be observed that for any $d$ the function $\pi_{d}$ can be
measured with a {\it single} collective entanglement witness as it
will be shown below, but it detects all the entanglement only in a
two-qubit case.

{\it Universal collective entanglement witness.-} Now we address a
natural question arising in the context of the results from the
previous section: {\it Is a measurement of a determinant of
$\varrho^{\Gamma}$ possible by means of a single observable?}
Following Ref. \cite{Witnesses} we define the collective witness
to be a Hermitian operator $W^{(n)}$, of which mean value on
$n$-copies of $\varrho$ is nonnegative, i.e,
$\srn{W^{(n)}}{\varrho^{\otimes
n}}:=\Tr\left(W^{(n)}\varrho^{\otimes n}\right) \geq 0$ and
negative on some entangled state.
Reformulating this question in terms of the above we ask if there
exists such an observable that
%
$\srn{W^{(4)}_{\mathrm{univ}}}{\varrho^{\otimes
n}}=\det\varrho^{\Gamma}.$
%
%
It has been shown \cite{Brun} that any $m$-th degree polynomial of
the elements of $\varrho$ (in particular its determinant) may be
found by determining an expectation value of two observables each
on $m$ copies of a state corresponding to real and imaginary part
of the value of the polynomial respectively. With guarantee ({\it
a priori} knowledge) that a polynomial is real valued we need only
single observable (cf. \cite{Witnesses}). In fact we deal with
such a polynomial here since the determinant (\ref{wyznacznik}) is
obviously real. It is a polynomial of the fourth degree  so the
necessary number of copies is four. This positively resolves the
problem of the existence of a single observable
$W^{(4)}_{\mathrm{univ}}$. To find the explicit form of it we
first introduce polynomials
%
$\Pi_{k}(\vec{x})= \sum_{i=1}^{m}x_{i}^{k},$
%
which for $\vec{x}=\vec{\lambda}$, a vector consisting of
eigenvalues of a given matrix, are just the $k$-th moments of this
matrix. We know that, for each $k$, $\Pi_{k}(\vec{\lambda})$ is
just a mean value of single observable
$\mathcal{O}^{(k)}=(1/2)\big(V^{(k)}+V^{(k)\dagger}\big)$ on $k$
copies of $\varrho$ with permutation operators $V^{(k)}$ defined
as $V^{(k)}\ket{\Phi_{1}}\ldots\ket{\Phi_{k-1}}\ket{\Phi_{k}}=
\ket{\Phi_{k}}\ket{\Phi_{1}}\ldots\ket{\Phi_{k-1}}$
$(k=1,\ldots,m)$, with $\ket{\Phi_{i}}\in\mathcal{H}$.

Now the crucial step is to connect the determinant of a matrix
with its easily measurable moments. Newton-Girard formulas
\cite{NewtonGirard} provide us with
%
$\det\varrho^{\Gamma}=(1/24)[1-6\Pi_4(\vec{\lambda})+8\Pi_3(\vec{\lambda})
+3\Pi_2^2(\vec{\lambda})-6\Pi_2(\vec{\lambda})].$
%
Before we proceed note that $V^{(k)}$ can be written in a
separable form as $\tilde{V}^{(k)}\ot \tilde{V}^{(k)}$ where
$\tilde{V}^{k}$ are permutations acting on the same subsystems of
 $\varrho^{\ot k}$.

The approach from Ref. \cite{Noiseless} leads to
\begin{eqnarray}
W^{(4)}_{\mathrm{univ}}=\frac{1}{24}\mathbbm{1}_{256}&-&\frac{1}{8}\big(\tilde{V}^{(4)}\ot
\tilde{V}^{(4)T}+\tilde{V}^{(4)T}\ot
\tilde{V}^{(4)}\big)\nonumber\\
&+&\frac{1}{6}\mathbbm{1}_{4}\ot
\big(\tilde{V}^{(3)}\ot\tilde{V}^{(3)T}+\tilde{V}^{(3)T}\ot
\tilde{V}^{(3)}\big)\nonumber\\
&+&\frac{1}{8}V^{(2)}\ot V^{(2)}-\frac{1}{4}\mathbbm{1}_{16}\ot
V^{(2)}
\end{eqnarray}
which mean value on four copies of $\varrho$ gives
$\det\varrho^{\Gamma}$.

{\it The network.-} Here we consider the problem of the
designation of a network measuring $W^{(4)}_{\mathrm{univ}}$.

The issue of avoiding unimportant data (frequency probabilities
corresponding to all eigenvalues of the observable) while
measuring the observable was considered in Refs.
\cite{Estimator,Paz}. The question about dimension of ancillas
involved in the measurement was answered in Ref. \cite{binpovm}
where it was shown that {\it via} unitary interaction with a
single qubit and final measurement of $\sigma_{z}$ on it, one can
get mean value of an arbitrary observable with bounded spectrum.
Finally, in Ref. \cite{Brun} it was shown that interaction between
systems in question and the ancilla can be conducted as a
controlled unitary operation. Note that the above single qubit
universality in a mean value estimation is compatible with the
further proof that single qubits are in a sense universal quantum
interfaces \cite{Interface}.

The most efficient in number of systems involved network involves
nine qubits interacting via unitary operation which can be
constructed in a way described in \cite{Noiseless}.

We present here (Fig.3) the alternative network that requires two
more ancillary qubits. However with this additional systems we
achieve simplicity of the structure of the controlled unitary
operations, which are just swaps. This device shows how one can
easily combine mean values of many observables. We do not go into
details concerning optimality of both networks in number of gates.
\begin{figure}\label{Fig3}
\centering
\includegraphics[width=8cm]{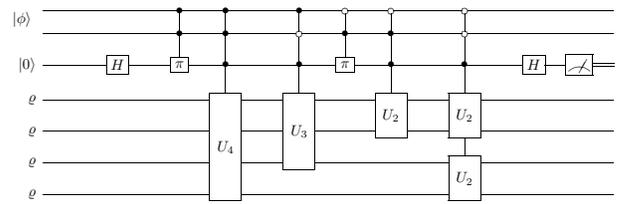}
\caption{Network determining entanglement properties of a two
qubit state by a single measurement of $\langle\sigma _z\rangle$
on a control qubit. Here
$\ket{\phi}=(1/\sqrt{23})(\sqrt{3}\ket{00}+\sqrt{6}
\ket{01}+\sqrt{8}\ket{10}+\sqrt{6}\ket{11})$, unitaries $U_i$ are
combinations of swap operations such that $\Tr U_i
\varrho^{\otimes i}=\Pi_i$. State would be declared entangled iff
the measurement yielded result less than $-1/23$.}
\end{figure}

{\it Generalizing the criterion.-} Here we discuss the above
approach in the context of entanglement of an arbitrary bipartite
state $\varrho$. Let $\Lambda$ be a positive, but not completely
positive, map. Following Ref. \cite{sep}, $\Lambda$ constitutes a
necessary separability condition for states acting on Hilbert
space $\mathcal{H}_{A}\ot\mathcal{H}_{B}$. One easily reformulates
this condition for separability in terms of determinant:

{\it Fact.-} If for given positive map $\Lambda$ it holds
$[I\ot\Lambda](\varrho)\geq 0$ then $\det\{[I \otimes
\Lambda](\varrho)\} \geq 0$.

In a general case the converse of the Fact fails which can be
shown by embedding entangled two-qubit state in a $3\otimes 3$
space. Note that, as shown in the Proposition, converse is true
for reduction applied to second subsystem of a $2 \otimes d$
system which is useful in context of entanglement distillability
(see \cite{xor}).

Construction of the proper observable along the lines of Ref.
\cite{Noiseless} results in an observable which mean value on $n$
copies of the state gives the desired determinant, i.e.,
$\srn{\tilde{W}_{\Lambda}^{(n)}}{\varrho^{\otimes n}}=
\det\{[I\ot\Lambda](\varrho)\}$.

The idea generalizes immediately to multiparty case where maps
positive on product states  \cite{MPRHmultiPLA} are involved.


{\it Conclusions.-} We have constructed {\it single} observable
test that detects entanglement of an {\it unknown} two-qubit
state. In addition, the function corresponding to it provides
bounds for negativity and concurrence. We have also designed the
quantum network that can also be interpreted as {\it a quantum
computing that solves quantitatively a problem with a quantum data
structure} (cf. \cite{PHAE}). \newline \indent Some research
towards higher dimensional generalizations has been initialized
however the results suffer from the lack of  character.
Nevertheless a very natural question arises: is there any other
way to generalize the main result, i.e., find single collective
observable that detects entanglement of any $d \otimes d $ quantum
system without ambiguity. For some SO(3)--invariant states
Proposition 1 was shown to hold in Ref. \cite{AugSta} thus giving
such an observable in case of these states. In general one would
first need some counterpart of the analytical criterion
(\ref{wyznacznik}) existence of  which is a long--standing open
problem in quantum information theory. The first question could be
whether there exists positive map which applied to one subsystem
of any bipartite density matrix produces full rank matrix with odd
number of negative eigenvalues so that the criterion based on the
determinant would remain true.

 {\it Acknowledgments .-} This work was prepared under the
Polish Ministry of Science and Education Grant No. 1 P03B 095 29
and EU IP Programme SCALA.


\begin{references}
\bibitem{QuEntEPR}M. A. Nielsen
and I. L. Chuang {\it "Quantum Computation and Quantum
Information"}, Cambridge University Press, Cambridge 2000; R. Horodecki {\it et al.}, quant-ph/0702225.




\bibitem{sep}M. Horodecki {\it et al.}, Phys. Lett. A  {\bf
223}, 1 (1996).

\bibitem{Terhal}
B. M. Terhal, Phys. Lett. A {\bf 271}, 319 (2000).

\bibitem{Developed}
M. Lewenstein {\it et al.}, Phys. Rev. A {\bf 62}, 052310 (2000); M.
Lewenstein {\it et al.}, {\it ibid.} {\bf 63}, 044304 (2001); A.
Sanpera {\it et al.}, {\it ibid.} {\bf 63}, 050301 (2001);
D. Bruss {\it et al.}, J. Mod. Opt. {\bf 49}, 1399 (2002); G.
T\'oth and O. G\"uhne, Phys. Rev. Lett. {\bf 94}, 060501 (2005);
F. Brand\~ao, Phys. Rev. A {\bf 72}, 022310 (2005).

\bibitem{Experiments}
M. Barbieri {\it et. al.}, Phys. Rev. Lett. {\bf 91}, 227901 (2003);
M. Bourennane {\it et al.}, {\it ibid.} {\bf 92}, 087902 (2004); K.
J. Resch {\it et al.}, {\it ibid.} {\bf 94}, 070402 (2005);
J. Altepeter et al., {\it ibid.} {\bf
95}, 033601 (2005); N. Kiesel {\it et. al.}, {\it ibid.} {\bf 95},
210502 (2005); H. H\"affner {\it et al.}, {\it ibid.} {\bf 438}, 643
(2005).

\bibitem{UncertaintyCV}
L. Duan {\it et. al.}, Phys. Rev. Lett. {\bf 84}, 2722 (2000); R.
Simon, {\it ibid.} {\bf 84}, 2726 (2000); N. Korolkova {\it et
al.}, Phys. Rev. A {\bf 65}, 052306 (2002); P. van Loock and A.
Furusawa, {\it ibid.} {\bf 67}, 052315 (2003); G. T\'oth {\it et
al.}, {\it ibid.} {\bf 68}, 062310 (2003);
E. Shchukin and W. Vogel, {\it ibid.} {\bf 95},
230502 (2005); P. Hyllus and J. Eisert, New J. Phys. {\bf 8}, 51
(2006).

\bibitem{UncertaintyDisc}
B. Julsgaard {\it et al.}, 
Nature {\bf 413}, 400 (2001); J. Uffink, Phys. Rev. Lett. {\bf
88}, 230406 (2002); A. Kuzmich, and E. S. Polzik, in {\it Quantum
Information with Continuous Variables}, edited by S. L. Braunstein
and A. K. Pati (Kluwer Academic Press, Dordrecht, 2003), p. 231;
S. Yu {\it et al.}, {\it ibid.} {\bf 91}, 217903 (2003); V.
Giovannetti {\it et al.}, Phys. Rev. A {\bf 67}, 022320 (2003);
H.F. Hofmann and S. Takeuchi, {\it ibid.} {\bf 68}, 032103 (2003);
O. G\"uhne, Phys. Rev. Lett. {\bf 92}, 117903 (2004); O. G\"uhne
and M. Lewenstein, Phys. Rev. A {\bf 70}, 022316 (2004); G. T\'oth
and O. G\"uhne, {\it ibid.} {\bf 72}, 022340 (2005); O. G\"uhne
and N. L\"utkenhaus, quant-ph/0512164; R. Augusiak {\it et al.},
Phys. Rev. A, in press (arXiv:0707.4315); R. Augusiak and J.
Stasi\'nska, arXiv:0709.3779.




\bibitem{PeresWotters}
A. Peres and W. K. Wootters, Phys. Rev. Lett. {\bf 66}, 1119
(1991).

\bibitem{SH00}
J. M. G.~Sancho and S. F. Huelga, Phys. Rev. A {\bf 61}, 042303
(2000).

\bibitem{Acin}
A. Acin {\it et al.},
Phys. Rev. A {\bf 61}, 062307 (2000).

\bibitem{PHAE}
P. Horodecki and A. Ekert, Phys. Rev. Lett {\bf 89}, 127902
(2002).


\bibitem{PHPRL}
P. Horodecki, Phys. Rev. Lett. {\bf 90}, 167901 (2003).

\bibitem{Carteret}
H. A. Carteret, Phys. Rev. Lett. {\bf 94}, 040502 (2005).

\bibitem{Noiseless}
P. Horodecki {\it et al.}, Phys. Rev. A {\bf 74}, 052323 (2006).

\bibitem{Estimator}
A. K. Ekert {\it et al.}, Phys. Rev. Lett. {\bf 88} 217901 (2002).

\bibitem{Filip} R. Filip, Phys. Rev. A {\bf 65}, 062320 (2002).


\bibitem{Fiurasek} J. Fiur\'{a}\v{s}ek, Phys. Rev. A {\bf 66}, 052315 (2002).

\bibitem{Witnesses}
P. Horodecki, Phys. Rev. A {\bf 68}, 052101 (2003).



\bibitem{NonlinExp}
F. A. Bovino {\it et al.}, Phys. Rev. Lett. {\bf 95}, 240407
(2005).

\bibitem{ConcurrenceDet}
S. P. Walborn {\it et al.},  Nature {\bf 440} (2006) 1022.

\bibitem{Mintert} F. Mintert and A. Buchleitner, quant-ph/0605250 (Phys. Rev. Lett. in press).

\bibitem{Magda} M. Stobi\'nska and K. W\'odkiewicz, Phys. Rev. A {\bf 71}, 032304 (2005).

\bibitem{Aolita} L. Aolita and F. Mintert, Phys. Rev. Lett. {\bf 97}, 50501
(2006).


\bibitem{MiKuBu2005}F. Mintert {\it et al.},
Phys. Rev. Lett. {\bf 95}, 260502 (2005).

\bibitem{Peres}
A. Peres, Phys. Rev. Lett. {\bf 77}, (1996) 1413.

\bibitem{Sanpera}
A. Sanpera {\it et al.}, 
Phys. Rev. A {\bf 58}, 826 (1999).

\bibitem{Verstraete}F. Verstraete {\it et al.}, J. Phys. A {\bf
34}, 1327 (2001).

\bibitem{Slater}P. B. Slater, Phys. Rev. A {\bf 71}, 052319 (2005).

\bibitem{VerstraeteSLOCC}
F. Verstraete {\it et al.},
Phys. Rev. A {\bf 68}, 012103 (2003).

\bibitem{Sinolecka}
M. Sino\l{}{\c{e}}cka {\it et al.}, Acta Phys. Pol. B, {\bf 33},
2081 (2002).

\bibitem{Gour} G. Gour, Phys. Rev. A, {\bf 71}, 012318 (2005).
\bibitem{xor}
M. Horodecki and P. Horodecki, Phys. Rev. A {\bf 59}, 4206 (1999).

\bibitem{red}N. J. Cerf {\it et al.}, Phys. Rev. A {\bf 60}, 898 (1999).

\bibitem{Negativity}G. Vidal and R. F. Werner, Phys. Rev. A, {\bf
65} 032314 (2002).



\bibitem{footnote}For a given pure state $\ket{\psi}$ concurrence is defined as
$C(\ket{\psi})=\sqrt{2(1-\Tr\varrho_{A}^{2})}$, where $\varrho_{A}=\Tr_{B}\proj{\psi}$, while
    for mixed state it is given by $C(\varrho)=\min\sum_{i}p_{i}C(\ket{\psi_{i}})$, where minimum is taken
    over all ensembles $\{p_{i},\ket{\psi_{i}}\}$ realizing $\varrho$ (W. K. Wootters,
    Quant. Inf. Comp., {\bf 1}, 27 (2001)). The negativity is defined as
    $N(\varrho)=||\varrho^{\Gamma}||_{1}-1$, where $\Gamma$ denotes the partial
    transposition (K. \.Zyczkowski {\it et al.},Phys. Rev. A {\bf 58}, 883 (1998)).

\bibitem{Verstraete2} F. Verstraete {\it et al.},
Phys. Rev. A {\bf 64}, 010101(R) (2001).


\bibitem{Belldiag} For singular states the subsequent argument can be easily modified
by considering non-singular perturbation of the state and applying
the continuity of $\pi_{2}$ and $C$.





\bibitem{MiranGrudka} A. Miranowicz and A. Grudka, Phys. Rev. A {\bf 70}, 032326
(2004); J. Opt. B {\bf 6}, 542 (2004).


\bibitem{Brun}
T. A. Brun, Quant. Inf. Comp. {\bf 4},  401 (2004);  M. Grassl
{\it et al.}, Phys. Rev. A {\bf 58}, 1833 (1998); M. S. Leifer
{\it et al.}, {\it ibid.} {\bf 69}, 052304 (2004).

\bibitem{NewtonGirard} R. Seroul, {\it Programming for
Mathematicians}, (Berlin, Springer 2000).


\bibitem{Paz}
J. P. Paz and A. Roncaglia, Phys. Rev. A {\bf 68}, 052316 (2003).

\bibitem{binpovm}
P. Horodecki, Phys. Rev. A {\bf 67}, 060101 (2003).

\bibitem{Interface}
S. Lloyd {\it et al.}, 
Phys. Rev. A {\bf 69}, 012305 (2004).

\bibitem{MPRHmultiPLA}M. Horodecki {\it et al.},
Phys. Lett. A {\bf 283}, 1 (2001).

\bibitem{AugSta} R. Augusiak and J. Stasi\'nska, Phys. Lett. A {\bf 363}, 3 (2007).



\end{references}
\end{document}